# Cyclocopula Technique to Study the Relationship Between Two Cyclostationary Time Series with Fractional Brownian Motion Errors

**Mohammadreza Mahmoudi [1], Amir Mosavi [2], ***

[1] Department of Statistics, Faculty of Science, Fasa University, Fasa, Fars, Iran
[2] Faculty of Engineering, Technische Universität Dresden, Dresden, Germany
amir.mosavi@mailbox.tu-dresden.de

**Abstract.** Detection of relationship between two time series is so important in environmental and hydrological studies. Several parametric and non-parametric approaches can be applied to detect relationships. These techniques are usually sensitive to stationarity assumption. In this research, a new copula- based method is introduced to detect the relationship between two cylostationary time series with fractional Brownian motion (fBm) errors. The numerical studies verify the performance of the introduced approach.

## 1. Introduction

The ways of modeling dependency between two time series has always been one of the main focuses in practice. The choice of the better model depends on the dependency structure of obligators and is crucial part of the modeling. The applied methods in previous studies include Pearson's correlation coefficient [1-6], Spearman's correlation coefficient [7-10], Kendall's correlation coefficient [11-14], Sen's slope [15-17], cross-correlation function [18-20] and copula [21-26]. When we face with the relationship of two stationary time series, cross-correlation function and copula are suggested. Cross-correlation function, as same as Pearson's correlation, is somewhat sensitive to abnormality of datasets and existence of outliers. In other words, for abnormal populations or when we face with outliers, cross-correlation function may not work well. An efficient way to model dependency is to use new modeling mechanism of Copula Theory which helps understand the correlation beyond linearity. In contrast, Copula is a function which transfers the multivariate distribution function to its marginal distribution function by quantiles. It is a great tool for modeling dependency where correlation follows the random distribution. Copula technique is most efficient for stationary time series and may not work well for non-stationary time series such as cyclostationary time series. To solve this issue, in this research, we introduce a copula-based regression technique. The ability of the proposed approach to detect relationship between two cyclostationary time series with fBm errors is studied. For this purpose, numerous datasets from two cyclostationary time series with fBm errors are produced and analyzed.

## 2. Methodology

### 2.1. Copula

Copula was first introduced by Sklar [27] as a statistical mechanism to transfer the joint distribution into its marginals and copula as a model to show the dependency between the marginals.

Copulas are functions link marginal distributions to the multivariate distributions which have well-defined properties [28]. Assume $X$ and $Y$ are two continuous random variables with a joint distribution function

$$H(x,y) = Pr(X \leq x, Y \leq y),$$

and marginals

$$F(x) = Pr(X \leq x),$$

and

$$G(y) = Pr(Y \leq y).$$

According to Sklar's theorem [27], there is a copula for all $x$ and $y$ in $[-\infty, \infty]$ satisfies the following equation:

$$H(x,y) = C(F(x), G(y)).$$

The theorem indicates that copula is joint distribution function, and joint distribution function can be also presented as copula given its marginal distributions. Thus, Schweizer [29] discussed that joint distribution modeling can be reduced to copula modeling. Since copula represents the variables' dependence, it's also named dependence function.

One of the main advantages of copulas is allowing to identify tail dependence across the multiple distributions. There are several Copulas that can be selected. The choice of most appropriate Copula has been an important issue. In practice, independent and perfectly correlated variables are naturally inapplicable for copulas. Due to its familiarity, Gaussian Copula [30] was the most famous one among others, however it failed to capture asymmetry, non-linearity and heavy tail dependency.

Therefore, alternative copulas have been used to model joint dependences. Several papers have been implied Copula families in modeling joint default dependencies in hydrology. Gumbel copula [31] is able to capture right tail dependence which is a particularly explored aspect of default dependency. Clayton copula [32] is defined to have left tail dependence. Student's $t$ copula [33] captures symmetric tail dependence with equally right and left tail dependence while Gaussian and Frank copulas [34] are defined as symmetric dependence without any tail dependence.

*2.2. Cyclostationary Time Series*

Stationarity is an important condition in classical time series analysis. But because of existence of cyclic rhythm in many practical situations such as hydrology and climatology, the stationarity assumption is not satisfied. As a suitable alternative, cyclostationary (CS) time series [35-36] are employed to describe the cyclic rhythms of rhythmic processes.

A time series $X_t$, is CS with cycle T (CS-T), if T is the minimum natural number so that

$$m(t) := E(X_t) = m(t + T),$$

and

$$R(s,t) := Cov(X_s, X_t) = E[(X_s - m(s))\overline{(X_t - m(t))}] = Cov(X_{s+T}, X_{t+T}),$$

for all integers s and t.

The cyclic correlation can be detected by employing the coherent statistics [36]. These statistics were defined by

$$|\hat{\gamma}(p,q,M)|^2 = \frac{\left|\sum_{m=0}^{M-1} d_X(\lambda_{p+m})\overline{d_X(\lambda_{q+m})}\right|^2}{\sum_{m=0}^{M-1}|d_X(\lambda_{p+m})|^2 \sum_{m=0}^{M-1}|d_X(\lambda_{q+m})|^2}, \quad p > 0, q \leq N,$$

where

$$d_X(\lambda) = n^{-1/2} \sum_{t=1}^{n} X_t e^{i(t-1)\lambda}, \lambda \in [0, 2\pi),$$

denotes the discrete Fourier transform (DFT) of $X_1, \ldots, X_n$, M refers to the the smoothness parameter, and

$$(\lambda_i, \lambda_j) \in \bigcup_{k \in \mathbb{Z}} \left\{ (\lambda_i, \lambda_j) \in [0, 2\pi)^2 : \lambda_j = \lambda_i + \frac{2\pi k}{T} \right\}.$$

Since the spectral domain of CS-T time series is supported on the lines

$$T_j(x) = x \pm \frac{2\pi(j-1)}{T}, j = 1, \ldots, T,$$

therefore, we expect the coherent statistics plot of a CS process emphasis these lines.

Because of cyclic rhythms in CS processes, using usual copula regression to compute the relationship of two CS-T time series is somewhat wrong. To solve this issue, in this research, a new approach is employed to evaluate relationship of two CS-T time series.

### 2.3. Fractional Brownian Motion

A fractional Brownian motion (fBm) with Hurst index $H \in (0,1)$, is defined by

$$B_H(t) = \frac{1}{\Gamma\left(H + \frac{1}{2}\right)} \int_0^t (t-s)^{H-\frac{1}{2}} dB(s),$$

where $B$ and $\Gamma$ are a Brownian motion process and gamma function [37-38], respectively. The auto-covariance function of the processes $B_H(t)$ and $B_H(s)$ are given by

$$\gamma(s,t) := Cov(B_H(s), B_H(t)) = \frac{1}{2}(|t|^{2H} + |s|^{2H} - |t-s|^{2H}).$$

## 2.4. Copula for Cyclostationary Time Series

As previously discussed, the copula technique is most efficient for stationary time series and may not work well for non-stationary time series such as cyclostationary time series. To solve this issue, in this research, we introduce a copula-based regression technique.

Assume $X_t$ and $Y_t$ are two CS-T time series. Let $\{x_1, \ldots, x_n\}$ and $\{y, \ldots, y_n\}$ $(n = mT, m \in N)$ are a path of $X_t$ and $Y_t$, respectively. The outline as our procedure as following:

(i) Split $\{x_1, \ldots, x_n\}$, $\{y_1, \ldots, y_n\}$ and $\{(x_1, y_1), \ldots, (x_n, y_n)\}$ into T partitions $\{x^{(1)}, \ldots, x^{(T)}\}$, $\{y^{(1)}, \ldots, y^{(T)},\}$ and $\{(x,y)^{(1)}, \ldots, (x,y)^{(T)}\}$, where

$$x^{(i)} = \{x_i, x_{i+T}, \ldots, x_{i+(m-1)T}\}, i = 1, \ldots, T,$$
$$y^{(i)} = \{y_i, y_{i+T}, \ldots, y_{i+(m-1)T}\}, i = 1, \ldots, T,$$

and

$$(x,y)^{(i)} = \{(x_i, y_i), (x_{i+T}, y_{i+T}), \ldots, (x_{i+(m-1)T}, y_{i+(m-1)T})\}, i = 1, \ldots, T.$$

(ii) Let $F_i$, $G_i$ and $H_i$ as the distribution function of the members of $x^{(i)}$, $y^{(i)}$, and $(x,y)^{(i)}$, respectively. Define the copula of $x^{(i)}$ and $y^{(i)}$ by

$$C_i = C(F_i, G_i), i = 1, \ldots, T,$$

and estimate the copula of $x^{(i)}$ and $y^{(i)}$ by

$$\hat{C}_i = \hat{C}(F_i, G_i), i = 1, \ldots, T.$$

(iii) Apply copula regression to find the regression equation of $y^{(i)}$ based on $x^{(i)}$,

$$\hat{y}_j = b_{0,i} + b_{0,i} x_j, i = 1, \ldots, T, j = i, i+T, i+(m-1)T.$$

(iv) Combine the regression equations to next uses such as prediction or goodness of fit tests.

**Remark 1:** If $X_t$ and $Y_t$ are respectively $CS - T_1$ and $CS - T_2$ time series, let $T = lcm(T_1, T_2)$, where lcm refers to smallest common multiple of $T_1$ and $T_2$.

## 3. Simulation

In this section, the ability of the proposed method to detect relationship between two CS time series is studied. For this purpose, numerous datasets from two CS time series $X_t$ and $Y_t$ are produced and analyzed.

The simulation procedure is as following:

Step 1: For fixed $n \in \{120, 240, 480, 1200\}$, $H \in \{0.25, 0.75\}$ and $T \in \{1, 2, 3, 4\}$, separate paths of size $n$ from two CS-T time series $X_t$ and $Y_t$ are produced. Numerous CS time series with different parameters are considered.

Step 2: The simulated dataset are split into T partitions $\{x^{(1)}, \ldots, x^{(T)}\}$ and $\{y^{(1)}, \ldots, y^{(T)}\}$. Then the copula of $x^{(i)}$ and $y^{(i)}$ is estimated by

$$\hat{C}_i = \hat{C}(F_i, G_i), i = 1, \ldots, T.$$

In this study, we apply five different copula families, namely Gaussian, Student's t copula and three Archimedean copulas; Clayton, Gumbel and Frank. We capture symmetric dependence without tail dependence with Gaussian and Frank copulas, symmetric dependence with upper and lower tail dependence with t copula, left (lower) tail dependence with Clayton copula, right (upper) tail dependence with Gumbel copula.

### 3.1. Gaussian Copula

Gaussian (normal) copula, as the name implies, assumes joint distribution follows bivariate standard normal distribution. Gaussian copula is the most commonly applied copula in practice due to its convenient properties. The bivariate Gaussian copula is presented by:

$$C(a, b) = M_n(\Phi^{-1}(a), \Phi^{-1}(b); \theta),$$

where $M_n$ is the joint bivariate cumulative standard normal distribution with $\theta \in [-1, 1]$, as the correlation of the bivariate normal distribution and $\Phi^{-1}$ is the inverse of a univariate standard normal distribution.

### 3.2. t Copula

The bivariate Student's $t$ copula is represented by:

$$C(a,b) = T_{\theta,v}\big(t_v^{-1}(a), t_v^{-1}(b)\big),$$

where $T_{R,v}$ is the standardized bivariate Student's $t$ distribution with covariance $\theta \in [-1,1]$ and $v$ degree of freedom. $t_v^{-1}(u_n)$ indicates the inverse of Student's $t$ cumulative distribution function. The main advantage of Student's $t$ copula over Gaussian copula is assuming a non-zero tail dependence even if correlation is zero.

### 3.3. Clayton copula

To estimate the lower tail dependency, the Clayton copula, is mostly suggested. The bivariate Clayton copula is represented by:

$$C(a,b) = \big(max(a^{-\theta} + b^{-\theta} - 1, 0)\big)^{-1/\theta},$$

where $\theta$ ($\theta \geq -1, \theta \neq 0$) is the copula. The parameter $\theta$ is related to Kendall's tau rank correlation $\tau$ as following:

$$\tau = \frac{\theta}{\theta+2}.$$

### 3.4. Gumbel copula

To capture weak lower tail dependence and strong upper tail dependence, the Gumbel copula is developed. The bivariate Gumbel copula is given by:

$$C(a,b) = \exp\Big(-\big((-\log a)^\theta + (-\log b)^\theta\big)^{\frac{1}{\theta}}\Big),$$

where $\theta \geq 1$ is the copula parameter. The Gumbel copula represents just independent and positive dependence.

The parameter $\theta$ is related to Kendall's tau rank correlation $\tau$ as following:

$$\tau = 1 - \theta^{-1}.$$

### 3.5. Frank copula

The bivariate Frank copula is given by:

$$C(a,b) = -\frac{1}{\theta}\log\left(1 + \frac{(exp(-\theta a)-1)(exp(-\theta b)-1)}{exp(-\theta)-1}\right),$$

where $\theta \neq 0$ is the copula parameter. Unlike the Gumbel and Clayton copulas, Frank copula allows both negative and positive dependence in data. The parameter $\theta$ is related to Kendall's tau rank correlation $\tau$ as following:

$$\tau = 1 + \frac{4[D_1(\theta)-1]}{\theta},$$

where

$$D_k(\alpha) = \frac{k}{\alpha^k}\int_0^\alpha \frac{t^k}{\exp(t)-1}dt, k=1,2.$$

Step 3: For each copula, the regression analysis is applied to estimate the equation of $y^{(i)}$ based on $x^{(i)}$,

$$\hat{y}_j = b_{0,i} + b_{0,i}x_j, i = 1, \ldots, T, j = i, i+T, i+(m-1)T.$$

Step 4: For each copula, the estimated copula regression equations are used to estimate T partitions $\{y^{(1)}, \ldots, y^{(T)}\}$ by $\{\hat{y}^{(1)}, \ldots, \hat{y}^{(T)}\}$.

Step 5: Different goodness of fit measures including Correlation coefficient (r), Willmott's Index (WI) and Nash-Sutcliffe coefficient (NS) are computed by

$$r = \frac{\sum_{i=1}^n (y_i - \bar{y})(\hat{y}_i - \bar{\hat{y}})}{\sqrt{\sum_{i=1}^n (y_i - \bar{y})^2}\sqrt{\sum_{i=1}^n (\hat{y}_i - \bar{\hat{y}})^2}},$$

$$WI = \frac{\sum_{i=1}^n (y_i - \hat{y}_i)^2}{\sum_{i=1}^n (|y_i - \bar{y}| + |\hat{y}_i - \bar{y}|)^2},$$

and

$$NS = 1 - \frac{\sum_{i=1}^n (y_i - \hat{y}_i)^2}{\sum_{i=1}^n (\hat{y}_i - \bar{\hat{y}})^2}.$$

Step 6: Steps 1 to 5 are repeated 1000 times.

Step 7: For each parameter setting, the means of $r, WI$ and $NS$ of all 1000 runs and five copulas are computed.

We consider first order periodic autoregressive with fBm error (PARFBM(1)) time series.

Assume the process

$$X_t = \phi(t)X_{t-1} + B_H,$$

and

$$Y_t = \alpha X_t + W_t, \quad \{W_t\} \sim IIDN(0,1),$$

where

$$\phi(t) = \frac{1+\phi\cos(2\pi t/T)}{2}.$$

Tables 1 and 2 summarize the computed values of goodness of fit indices, for different parameter settings and $H = 0.25$ and $H = 0.75$, respectively. The results show that the values of goodness of fit indices are close to one. In other words, proposed method is robust to detect relationship between two cyclostationary time series.

Table 1: Testing performance of the Periodic Copula Model to detect relationship between two PARFBM(1) time series ($H = 0.25$), in terms of the Correlation Coefficient (r), Nash-Sutcliff (NSE) and Willmott Index (WI)

| Copula | T | $\phi$ | $\alpha$ | N | | | | | | | | | | | |
|---|---|---|---|---|---|---|---|---|---|---|---|---|---|---|---|
| | | | | 120 | | | 240 | | | 480 | | | 1200 | | |
| | | | | r | WI | NS | r | WI | NS | r | WI | NS | r | WI | NS |
| Gaussian | 1 | 0.3 | 0.3 | 0.980 | 0.992 | 0.989 | 0.978 | 0.991 | 0.988 | 0.978 | 0.983 | 0.990 | 0.998 | 0.995 | 0.980 |
| | | 0.3 | 0.7 | 0.977 | 0.989 | 0.998 | 0.985 | 0.992 | 0.992 | 0.970 | 0.978 | 0.998 | 0.983 | 0.997 | 0.997 |
| | | 0.7 | 0.3 | 0.994 | 0.985 | 0.999 | 0.981 | 0.990 | 0.973 | 0.996 | 0.977 | 0.996 | 0.989 | 0.994 | 0.990 |
| | | 0.7 | 0.7 | 0.984 | 0.989 | 0.991 | 0.988 | 0.998 | 0.987 | 0.983 | 0.986 | 0.985 | 0.983 | 0.986 | 0.976 |
| | 2 | 0.3 | 0.3 | 0.992 | 0.987 | 0.973 | 0.994 | 1.000 | 0.976 | 0.987 | 0.983 | 0.994 | 0.977 | 0.993 | 0.987 |
| | | 0.3 | 0.7 | 0.984 | 0.976 | 0.995 | 0.984 | 0.991 | 0.995 | 0.987 | 0.990 | 0.984 | 0.973 | 0.992 | 0.972 |
| | | 0.7 | 0.3 | 0.997 | 0.986 | 0.999 | 0.998 | 0.981 | 0.989 | 0.992 | 0.987 | 0.986 | 0.987 | 0.975 | 0.990 |
| | | 0.7 | 0.7 | 0.970 | 0.972 | 0.978 | 0.988 | 0.992 | 0.979 | 0.982 | 0.993 | 0.985 | 0.971 | 0.986 | 0.978 |
| | 3 | 0.3 | 0.3 | 0.978 | 0.983 | 0.982 | 0.996 | 0.983 | 0.982 | 0.979 | 0.976 | 0.976 | 0.996 | 0.994 | 0.987 |
| | | 0.3 | 0.7 | 0.970 | 0.984 | 0.975 | 0.981 | 0.992 | 0.975 | 0.975 | 1.000 | 0.987 | 0.983 | 0.982 | 0.971 |
| | | 0.7 | 0.3 | 0.981 | 0.992 | 0.997 | 0.976 | 0.980 | 0.989 | 0.980 | 0.994 | 0.989 | 0.977 | 0.984 | 0.981 |
| | | 0.7 | 0.7 | 0.979 | 0.971 | 0.982 | 0.992 | 0.991 | 0.989 | 0.998 | 0.995 | 0.990 | 0.980 | 0.999 | 0.992 |
| | 4 | 0.3 | 0.3 | 0.992 | 0.986 | 0.972 | 0.985 | 0.975 | 0.984 | 0.986 | 0.970 | 0.980 | 0.997 | 0.971 | 0.990 |
| | | 0.3 | 0.7 | 0.983 | 0.982 | 0.970 | 0.972 | 0.994 | 0.978 | 0.985 | 0.983 | 0.978 | 0.970 | 0.996 | 0.972 |
| | | 0.7 | 0.3 | 0.985 | 0.977 | 0.978 | 0.982 | 0.977 | 0.982 | 0.987 | 0.994 | 0.987 | 0.981 | 0.993 | 0.995 |
| | | 0.7 | 0.7 | 0.989 | 0.989 | 0.980 | 0.977 | 0.972 | 0.983 | 0.972 | 0.989 | 0.996 | 0.993 | 0.972 | 0.997 |
| | 1 | 0.3 | 0.3 | 0.996 | 0.971 | 0.996 | 0.978 | 0.971 | 0.996 | 0.991 | 0.978 | 0.986 | 0.998 | 0.979 | 0.975 |
| T | | 0.3 | 0.7 | 1.000 | 0.978 | 0.998 | 0.996 | 0.979 | 0.991 | 0.999 | 0.984 | 0.989 | 0.973 | 0.981 | 0.997 |
| | | 0.7 | 0.3 | 0.974 | 0.996 | 0.990 | 0.979 | 0.986 | 0.976 | 0.993 | 0.986 | 0.974 | 0.988 | 0.975 | 0.976 |
| | | 0.7 | 0.7 | 0.985 | 0.985 | 0.999 | 0.984 | 0.976 | 0.972 | 0.979 | 0.998 | 0.973 | 0.988 | 0.975 | 0.983 |
| | 2 | 0.3 | 0.3 | 0.978 | 0.971 | 0.985 | 0.983 | 0.997 | 0.973 | 0.976 | 0.973 | 0.985 | 0.994 | 0.998 | 0.979 |
| | | 0.3 | 0.7 | 0.994 | 0.990 | 0.986 | 0.984 | 0.975 | 0.999 | 0.994 | 0.987 | 0.973 | 0.984 | 0.995 | 0.999 |
| | | 0.7 | 0.3 | 0.983 | 0.970 | 0.997 | 0.981 | 0.983 | 0.999 | 0.993 | 0.972 | 0.987 | 0.971 | 0.970 | 0.995 |
| | | 0.7 | 0.7 | 0.984 | 0.991 | 0.978 | 0.972 | 0.988 | 0.992 | 0.983 | 0.973 | 0.991 | 0.998 | 0.994 | 0.980 |
| | 3 | 0.3 | 0.3 | 0.993 | 0.979 | 0.978 | 0.983 | 0.992 | 0.993 | 0.979 | 0.978 | 0.985 | 0.977 | 0.998 | 0.971 |
| | | 0.3 | 0.7 | 0.979 | 0.993 | 0.996 | 0.989 | 0.993 | 0.995 | 0.989 | 0.999 | 0.996 | 0.971 | 0.980 | 0.984 |
| | | 0.7 | 0.3 | 0.999 | 0.975 | 0.982 | 0.979 | 0.994 | 0.996 | 0.977 | 0.996 | 1.000 | 0.997 | 0.993 | 1.000 |
| | | 0.7 | 0.7 | 0.986 | 0.986 | 0.991 | 0.970 | 0.992 | 0.978 | 0.978 | 0.992 | 0.997 | 0.976 | 0.986 | 0.978 |
| | 4 | 0.3 | 0.3 | 0.997 | 0.976 | 0.979 | 0.970 | 0.992 | 0.976 | 0.974 | 0.970 | 0.980 | 0.998 | 0.999 | 0.979 |
| | | 0.3 | 0.7 | 0.977 | 0.975 | 0.979 | 0.994 | 0.989 | 0.980 | 1.000 | 0.994 | 0.974 | 0.970 | 0.984 | 0.982 |
| | | 0.7 | 0.3 | 0.986 | 0.983 | 0.984 | 0.992 | 0.980 | 0.976 | 0.978 | 0.987 | 0.981 | 0.973 | 0.982 | 0.986 |
| | | 0.7 | 0.7 | 0.993 | 0.971 | 0.992 | 0.998 | 0.996 | 0.999 | 0.992 | 0.984 | 0.998 | 0.996 | 0.976 | 0.982 |
| Clayton | 1 | 0.3 | 0.3 | 1.000 | 0.979 | 0.985 | 0.977 | 0.987 | 0.984 | 0.980 | 0.991 | 0.986 | 0.991 | 0.987 | 0.986 |
| | | 0.3 | 0.7 | 0.981 | 0.999 | 0.999 | 0.996 | 0.996 | 0.998 | 0.992 | 0.993 | 0.981 | 0.984 | 0.986 | 0.989 |
| | | 0.7 | 0.3 | 0.973 | 0.995 | 0.997 | 0.992 | 0.979 | 0.990 | 0.971 | 0.996 | 0.990 | 0.988 | 0.972 | 0.980 |
| | | 0.7 | 0.7 | 0.986 | 0.986 | 0.979 | 0.994 | 0.977 | 0.996 | 0.989 | 0.998 | 0.983 | 0.992 | 0.976 | 0.982 |
| | 2 | 0.3 | 0.3 | 0.982 | 0.996 | 0.993 | 0.988 | 0.975 | 0.990 | 0.986 | 0.983 | 0.980 | 0.972 | 0.997 | 0.990 |
| | | 0.3 | 0.7 | 0.984 | 0.975 | 0.981 | 0.986 | 0.983 | 0.985 | 0.977 | 0.970 | 0.974 | 0.975 | 0.999 | 0.982 |
| | | 0.7 | 0.3 | 0.989 | 0.992 | 0.978 | 0.981 | 0.971 | 0.995 | 0.995 | 0.975 | 0.994 | 0.982 | 0.971 | 0.987 |
| | | 0.7 | 0.7 | 0.993 | 0.988 | 0.976 | 0.982 | 0.991 | 0.987 | 0.996 | 0.986 | 0.987 | 0.997 | 0.973 | 0.970 |
| | 3 | 0.3 | 0.3 | 0.988 | 0.978 | 0.975 | 0.987 | 0.989 | 0.994 | 0.995 | 0.982 | 0.972 | 0.984 | 0.989 | 0.983 |
| | | 0.3 | 0.7 | 0.973 | 0.998 | 0.980 | 0.980 | 0.980 | 0.988 | 0.982 | 0.981 | 0.983 | 0.993 | 0.975 | 0.991 |
| | | 0.7 | 0.3 | 0.995 | 0.999 | 0.989 | 0.992 | 0.985 | 0.993 | 0.999 | 0.990 | 0.979 | 0.979 | 0.990 | 0.978 |
| | | 0.7 | 0.7 | 1.000 | 0.991 | 0.971 | 0.992 | 0.998 | 0.992 | 0.972 | 0.981 | 0.978 | 0.988 | 0.987 | 0.971 |
| | 4 | 0.3 | 0.3 | 0.981 | 0.989 | 0.983 | 0.975 | 0.997 | 0.995 | 0.989 | 0.991 | 0.981 | 0.993 | 0.996 | 0.978 |
| | | 0.3 | 0.7 | 0.975 | 0.984 | 0.971 | 0.987 | 0.983 | 0.970 | 0.972 | 1.000 | 0.981 | 0.989 | 0.987 | 1.000 |
| | | 0.7 | 0.3 | 0.971 | 0.988 | 0.981 | 0.975 | 0.971 | 0.999 | 0.996 | 0.977 | 1.000 | 0.991 | 0.997 | 0.992 |
| | | 0.7 | 0.7 | 0.971 | 0.996 | 0.980 | 1.000 | 0.972 | 0.977 | 0.987 | 0.973 | 0.975 | 0.993 | 0.985 | 0.998 |
| Gumbel | 1 | 0.3 | 0.3 | 0.997 | 0.990 | 0.983 | 0.979 | 0.985 | 0.982 | 0.987 | 0.983 | 0.997 | 0.979 | 0.993 | 0.983 |
| | | 0.3 | 0.7 | 0.972 | 0.974 | 0.984 | 0.971 | 0.972 | 0.997 | 0.989 | 0.983 | 0.997 | 0.978 | 0.997 | 0.985 |
| | | 0.7 | 0.3 | 0.988 | 0.974 | 0.999 | 0.979 | 0.996 | 0.974 | 0.979 | 0.985 | 0.973 | 0.987 | 0.980 | 0.993 |
| | | 0.7 | 0.7 | 0.987 | 0.987 | 0.992 | 0.988 | 0.998 | 0.993 | 0.975 | 0.973 | 0.978 | 0.970 | 0.987 | 0.990 |
| | 2 | 0.3 | 0.3 | 0.996 | 0.996 | 0.988 | 0.991 | 0.980 | 0.985 | 0.986 | 0.973 | 0.988 | 0.996 | 0.995 | 0.997 |
| | | 0.3 | 0.7 | 0.998 | 0.998 | 0.973 | 1.000 | 0.983 | 0.972 | 0.996 | 0.981 | 0.986 | 0.973 | 0.976 | 0.998 |
| | | 0.7 | 0.3 | 0.971 | 0.979 | 0.974 | 0.982 | 0.997 | 0.992 | 0.974 | 0.995 | 0.990 | 0.974 | 0.977 | 0.986 |
| | | 0.7 | 0.7 | 0.975 | 0.993 | 0.995 | 0.980 | 0.992 | 0.987 | 0.994 | 0.984 | 0.972 | 0.997 | 0.970 | 0.971 |
| | 3 | 0.3 | 0.3 | 0.977 | 0.979 | 0.977 | 0.995 | 0.996 | 0.993 | 0.978 | 0.975 | 0.997 | 0.987 | 0.982 | 0.999 |
| | | 0.3 | 0.7 | 0.983 | 0.999 | 0.975 | 0.995 | 0.984 | 0.992 | 0.993 | 0.971 | 0.974 | 0.998 | 0.970 | 0.972 |
| | | 0.7 | 0.3 | 0.989 | 0.972 | 0.988 | 0.970 | 0.976 | 0.999 | 0.993 | 0.972 | 0.982 | 0.996 | 0.981 | 0.986 |
| | | 0.7 | 0.7 | 0.983 | 0.999 | 0.977 | 0.993 | 0.973 | 0.975 | 0.972 | 0.971 | 0.990 | 0.998 | 0.979 | 0.985 |
| | 4 | 0.3 | 0.3 | 0.970 | 0.982 | 0.984 | 0.974 | 0.974 | 0.973 | 0.975 | 0.974 | 0.974 | 0.977 | 0.981 | 0.993 |
| | | 0.3 | 0.7 | 0.992 | 0.974 | 0.974 | 0.971 | 0.994 | 0.982 | 0.986 | 0.981 | 0.996 | 0.976 | 0.986 | 0.984 |
| | | 0.7 | 0.3 | 0.996 | 0.973 | 0.977 | 0.972 | 0.974 | 0.991 | 0.999 | 0.981 | 0.988 | 0.973 | 0.998 | 0.973 |
| | | 0.7 | 0.7 | 0.981 | 0.981 | 0.984 | 0.995 | 0.990 | 0.991 | 0.973 | 0.979 | 0.976 | 0.983 | 0.978 | 0.979 |
| Frank | 1 | 0.3 | 0.3 | 0.983 | 0.997 | 0.989 | 0.987 | 0.999 | 0.985 | 0.984 | 0.972 | 0.989 | 0.999 | 0.980 | 0.984 |
| | | 0.3 | 0.7 | 0.975 | 0.976 | 0.974 | 0.979 | 0.995 | 0.981 | 0.980 | 0.979 | 0.991 | 0.972 | 0.970 | 0.997 |
| | | 0.7 | 0.3 | 0.982 | 1.000 | 0.995 | 0.972 | 0.999 | 0.978 | 0.974 | 0.999 | 0.974 | 0.986 | 0.974 | 0.992 |
| | | 0.7 | 0.7 | 0.998 | 0.973 | 0.973 | 0.973 | 0.994 | 0.993 | 0.992 | 0.993 | 0.982 | 0.987 | 0.991 | 0.976 |
| | 2 | 0.3 | 0.3 | 0.997 | 0.971 | 0.972 | 0.972 | 0.976 | 0.992 | 0.988 | 0.981 | 0.971 | 0.994 | 0.995 | 0.986 |
| | | 0.3 | 0.7 | 0.998 | 0.985 | 0.970 | 0.982 | 0.998 | 0.971 | 0.998 | 0.978 | 0.985 | 0.976 | 0.993 | 0.999 |
| | | 0.7 | 0.3 | 0.979 | 0.984 | 0.991 | 0.990 | 0.975 | 0.999 | 0.970 | 0.977 | 0.999 | 0.973 | 0.983 | 0.986 |
| | | 0.7 | 0.7 | 0.983 | 0.995 | 0.993 | 0.973 | 0.989 | 0.994 | 0.999 | 1.000 | 0.971 | 0.986 | 0.979 | 0.993 |
| | 3 | 0.3 | 0.3 | 0.994 | 0.988 | 0.978 | 0.984 | 0.978 | 0.982 | 0.996 | 0.978 | 0.971 | 0.997 | 0.975 | 0.992 |
| | | 0.3 | 0.7 | 0.991 | 0.990 | 0.984 | 0.989 | 0.976 | 0.987 | 0.972 | 0.973 | 0.994 | 0.976 | 0.985 | 0.994 |
| | | 0.7 | 0.3 | 0.994 | 0.977 | 0.996 | 1.000 | 0.978 | 0.995 | 0.976 | 0.996 | 0.976 | 0.983 | 0.995 | 0.990 |
| | | 0.7 | 0.7 | 0.980 | 0.978 | 0.971 | 0.989 | 0.979 | 0.991 | 0.980 | 0.988 | 0.993 | 0.972 | 0.987 | 0.986 |

| | | | | 4 | 0.3 | 0.3 | 0.981 | 0.971 | 0.988 | 0.979 | 0.990 | 0.981 | 0.990 | 0.987 | 0.972 | 0.994 | 0.982 | 0.996 |
| | | | | | 0.3 | 0.7 | 0.988 | 0.987 | 0.973 | 0.976 | 0.980 | 0.992 | 0.976 | 0.990 | 0.985 | 0.976 | 0.992 | 0.978 |
| | | | | | 0.7 | 0.3 | 0.975 | 0.979 | 0.970 | 0.997 | 0.975 | 0.994 | 0.978 | 0.990 | 0.977 | 0.994 | 0.971 | 0.999 |
| | | | | | 0.7 | 0.7 | 0.986 | 0.985 | 0.990 | 0.994 | 0.979 | 0.984 | 0.976 | 0.998 | 0.987 | 0.990 | 0.996 | 0.983 |

Table 2: Testing performance of the Periodic Copula Model to detect relationship between two PARFBM(1) time series ($H = 0.25$), in terms of the Correlation Coefficient (r), Nash-Sutcliff (NSE) and Willmott Index (WI)

| Copula | T | θ | β | n | | | | | | | | | | | | |
|---|---|---|---|---|---|---|---|---|---|---|---|---|---|---|---|---|
| | | | | 120 | | | 240 | | | 480 | | | 1200 | | |
| | | | | r | WI | NS | r | WI | NS | r | WI | NS | r | WI | NS |
| Gaussian | 1 | 0.3 | 0.3 | 0.990 | 0.996 | 0.981 | 0.982 | 1.000 | 0.980 | 0.986 | 0.998 | 0.989 | 0.977 | 0.992 | 0.994 |
| | | 0.3 | 0.7 | 0.975 | 0.978 | 0.981 | 0.997 | 0.992 | 0.990 | 0.971 | 0.981 | 0.996 | 0.990 | 0.971 | 0.978 |
| | | 0.7 | 0.3 | 0.990 | 0.997 | 0.984 | 0.995 | 0.997 | 0.973 | 0.974 | 0.999 | 0.992 | 0.983 | 0.985 | 0.974 |
| | | 0.7 | 0.7 | 0.995 | 0.999 | 0.983 | 0.996 | 0.994 | 0.974 | 0.998 | 0.986 | 0.972 | 0.977 | 0.971 | 0.996 |
| | 2 | 0.3 | 0.3 | 0.985 | 0.985 | 0.997 | 0.992 | 0.979 | 0.992 | 0.999 | 0.978 | 0.985 | 0.999 | 0.971 | 0.980 |
| | | 0.3 | 0.7 | 0.971 | 0.999 | 0.988 | 0.995 | 0.996 | 0.983 | 0.993 | 0.975 | 0.991 | 0.970 | 0.988 | 0.983 |
| | | 0.7 | 0.3 | 0.999 | 0.971 | 0.983 | 0.979 | 0.993 | 0.994 | 0.984 | 0.988 | 0.972 | 0.976 | 0.983 | 0.991 |
| | | 0.7 | 0.7 | 0.994 | 0.985 | 0.993 | 0.971 | 0.979 | 0.976 | 0.971 | 0.985 | 0.989 | 0.996 | 0.994 | 0.987 |
| | 3 | 0.3 | 0.3 | 0.992 | 0.993 | 0.992 | 0.971 | 1.000 | 0.991 | 0.995 | 0.985 | 0.994 | 0.988 | 0.982 | 0.994 |
| | | 0.3 | 0.7 | 0.996 | 0.997 | 0.977 | 0.981 | 0.991 | 0.975 | 0.991 | 0.980 | 0.987 | 0.995 | 1.000 | 0.982 |
| | | 0.7 | 0.3 | 0.998 | 0.976 | 0.980 | 0.974 | 0.974 | 0.982 | 0.984 | 0.979 | 0.996 | 0.993 | 0.971 | 0.975 |
| | | 0.7 | 0.7 | 0.988 | 0.985 | 0.983 | 0.976 | 0.993 | 0.993 | 0.985 | 0.985 | 0.995 | 0.989 | 0.994 | 0.978 |
| | 4 | 0.3 | 0.3 | 0.995 | 0.996 | 0.990 | 0.991 | 0.989 | 0.996 | 0.999 | 0.990 | 0.978 | 0.993 | 0.987 | 0.980 |
| | | 0.3 | 0.7 | 1.000 | 0.988 | 0.990 | 0.993 | 0.970 | 0.972 | 0.993 | 0.982 | 0.977 | 0.999 | 0.985 | 0.972 |
| | | 0.7 | 0.3 | 0.983 | 0.980 | 1.000 | 0.993 | 0.977 | 0.972 | 0.998 | 0.978 | 0.998 | 0.973 | 0.970 | 0.993 |
| | | 0.7 | 0.7 | 0.991 | 0.977 | 0.985 | 0.994 | 0.983 | 0.973 | 0.983 | 0.975 | 0.990 | 0.983 | 0.982 | 0.990 |
| T | 1 | 0.3 | 0.3 | 0.990 | 0.996 | 0.981 | 0.974 | 0.971 | 0.998 | 0.992 | 0.989 | 0.982 | 0.972 | 0.980 | 0.974 |
| | | 0.3 | 0.7 | 0.972 | 0.987 | 0.990 | 0.993 | 0.996 | 0.998 | 0.986 | 0.972 | 0.994 | 0.992 | 0.978 | 0.975 |
| | | 0.7 | 0.3 | 0.977 | 0.984 | 0.997 | 0.974 | 0.999 | 0.996 | 0.996 | 0.989 | 0.986 | 0.992 | 0.995 | 0.980 |
| | | 0.7 | 0.7 | 0.973 | 0.989 | 0.974 | 0.981 | 0.973 | 0.982 | 0.992 | 0.979 | 0.985 | 0.994 | 0.980 | 0.973 |
| | 2 | 0.3 | 0.3 | 0.996 | 0.975 | 0.999 | 0.976 | 0.995 | 0.979 | 0.978 | 0.985 | 0.979 | 0.984 | 0.999 | 0.990 |
| | | 0.3 | 0.7 | 0.977 | 0.979 | 0.995 | 0.995 | 0.976 | 0.981 | 0.973 | 0.974 | 0.983 | 0.989 | 0.976 | 0.979 |
| | | 0.7 | 0.3 | 0.996 | 0.971 | 0.977 | 0.983 | 0.981 | 0.989 | 0.972 | 0.998 | 0.984 | 0.980 | 0.990 | 0.981 |
| | | 0.7 | 0.7 | 0.994 | 0.990 | 0.997 | 0.997 | 0.992 | 0.980 | 0.987 | 0.978 | 0.987 | 0.992 | 0.986 | 0.974 |
| | 3 | 0.3 | 0.3 | 0.980 | 0.986 | 0.992 | 0.979 | 0.973 | 0.998 | 0.997 | 0.988 | 0.975 | 0.988 | 0.981 | 0.992 |
| | | 0.3 | 0.7 | 0.978 | 0.999 | 0.984 | 0.990 | 0.986 | 0.993 | 0.978 | 0.980 | 0.977 | 0.972 | 0.979 | 0.987 |
| | | 0.7 | 0.3 | 0.983 | 0.986 | 0.983 | 0.993 | 0.986 | 0.987 | 0.971 | 0.992 | 0.980 | 0.972 | 0.985 | 0.979 |
| | | 0.7 | 0.7 | 0.998 | 0.988 | 0.991 | 0.987 | 0.984 | 0.977 | 0.980 | 0.993 | 0.997 | 0.985 | 0.989 | 0.985 |
| | 4 | 0.3 | 0.3 | 0.980 | 1.000 | 0.994 | 0.990 | 0.988 | 0.982 | 0.980 | 0.986 | 0.997 | 0.996 | 0.987 | 0.973 |
| | | 0.3 | 0.7 | 0.989 | 0.978 | 0.984 | 0.987 | 0.979 | 0.976 | 0.985 | 0.995 | 0.989 | 0.972 | 0.974 | 0.986 |
| | | 0.7 | 0.3 | 0.975 | 0.978 | 0.982 | 0.982 | 0.990 | 0.993 | 0.977 | 0.981 | 0.972 | 0.985 | 0.999 | 0.993 |
| | | 0.7 | 0.7 | 0.985 | 0.995 | 0.974 | 0.986 | 0.995 | 0.993 | 0.972 | 0.990 | 0.994 | 0.996 | 0.991 | 1.000 |
| | 1 | 0.3 | 0.3 | 0.986 | 0.985 | 0.998 | 0.980 | 0.974 | 0.978 | 0.992 | 0.978 | 1.000 | 0.971 | 0.984 | 0.975 |
| | | 0.3 | 0.7 | 0.986 | 0.973 | 0.977 | 0.984 | 0.980 | 0.993 | 0.986 | 0.971 | 0.999 | 0.979 | 0.980 | 0.980 |
| | | 0.7 | 0.3 | 0.995 | 0.979 | 0.987 | 0.975 | 0.977 | 0.972 | 0.981 | 0.994 | 0.975 | 0.982 | 0.977 | 0.980 |
| | | 0.7 | 0.7 | 0.995 | 0.984 | 0.974 | 0.973 | 0.995 | 0.975 | 0.987 | 0.974 | 0.987 | 0.993 | 0.976 | 0.978 |
| | 2 | 0.3 | 0.3 | 0.987 | 0.999 | 0.983 | 0.991 | 0.974 | 0.989 | 0.979 | 0.972 | 0.996 | 0.994 | 0.996 | 0.981 |
| | | 0.3 | 0.7 | 0.981 | 0.980 | 0.973 | 0.989 | 0.975 | 0.995 | 0.995 | 0.976 | 0.974 | 0.981 | 0.980 | 1.000 |
| | | 0.7 | 0.3 | 0.978 | 1.000 | 0.977 | 0.970 | 0.994 | 0.993 | 0.985 | 0.971 | 0.975 | 0.974 | 0.978 | 0.989 |
| | | 0.7 | 0.7 | 0.992 | 0.980 | 0.972 | 0.973 | 0.974 | 0.981 | 0.973 | 0.980 | 0.975 | 0.994 | 0.986 | 0.975 |

| Family | θ | u₁ | u₂ | | | | | | | | | | | | |
|---|---|---|---|---|---|---|---|---|---|---|---|---|---|---|---|
| Clayton | 3 | 0.3 | 0.3 | 0.995 | 0.987 | 0.974 | 0.988 | 0.989 | 0.989 | 0.986 | 0.976 | 0.992 | 0.992 | 0.991 | 0.981 |
| | | 0.3 | 0.7 | 0.972 | 0.987 | 0.999 | 0.984 | 0.988 | 0.991 | 0.991 | 0.974 | 0.987 | 0.978 | 0.981 | 0.977 |
| | | 0.7 | 0.3 | 0.997 | 0.985 | 0.998 | 0.987 | 0.978 | 0.973 | 0.979 | 0.972 | 0.983 | 0.994 | 0.977 | 0.974 |
| | | 0.7 | 0.7 | 0.996 | 0.989 | 0.999 | 0.998 | 0.980 | 0.996 | 0.975 | 0.970 | 0.983 | 0.996 | 0.976 | 0.986 |
| | 4 | 0.3 | 0.3 | 0.978 | 0.981 | 0.987 | 0.980 | 0.979 | 0.988 | 0.983 | 0.989 | 0.974 | 0.975 | 0.991 | 0.992 |
| | | 0.3 | 0.7 | 0.994 | 0.976 | 0.993 | 0.970 | 0.983 | 0.999 | 0.980 | 0.976 | 0.986 | 0.975 | 0.981 | 0.992 |
| | | 0.7 | 0.3 | 0.991 | 0.999 | 0.977 | 0.981 | 0.979 | 0.992 | 0.976 | 0.971 | 0.989 | 0.995 | 0.978 | 0.971 |
| | | 0.7 | 0.7 | 0.978 | 0.977 | 0.981 | 0.976 | 0.988 | 0.989 | 0.979 | 0.977 | 0.979 | 0.985 | 0.985 | 0.988 |
| Gumbel | 1 | 0.3 | 0.3 | 0.981 | 0.986 | 0.982 | 0.973 | 0.971 | 0.970 | 0.982 | 0.999 | 0.986 | 0.994 | 0.983 | 0.999 |
| | | 0.3 | 0.7 | 0.991 | 1.000 | 0.991 | 0.977 | 0.980 | 0.987 | 0.976 | 0.996 | 0.993 | 0.979 | 0.990 | 0.975 |
| | | 0.7 | 0.3 | 0.997 | 0.971 | 0.975 | 0.990 | 0.986 | 0.994 | 0.972 | 0.983 | 0.996 | 0.994 | 0.972 | 0.994 |
| | | 0.7 | 0.7 | 0.987 | 0.988 | 0.993 | 0.977 | 0.993 | 0.994 | 0.997 | 0.995 | 0.977 | 0.977 | 0.981 | 0.982 |
| | 2 | 0.3 | 0.3 | 0.983 | 0.982 | 0.983 | 0.971 | 0.985 | 0.979 | 0.979 | 0.985 | 0.972 | 0.975 | 1.000 | 0.977 |
| | | 0.3 | 0.7 | 0.997 | 0.975 | 0.970 | 0.996 | 0.987 | 0.976 | 0.983 | 0.981 | 0.982 | 0.995 | 0.980 | 0.973 |
| | | 0.7 | 0.3 | 0.985 | 0.990 | 0.974 | 0.999 | 0.989 | 0.996 | 0.999 | 0.994 | 0.989 | 0.985 | 0.983 | 0.971 |
| | | 0.7 | 0.7 | 0.994 | 0.995 | 0.999 | 0.999 | 0.993 | 0.981 | 0.994 | 0.985 | 0.970 | 0.996 | 0.990 | 0.994 |
| | 3 | 0.3 | 0.3 | 0.972 | 0.979 | 0.985 | 0.974 | 1.000 | 0.974 | 0.976 | 0.992 | 0.995 | 0.981 | 0.982 | 0.986 |
| | | 0.3 | 0.7 | 0.975 | 0.977 | 0.985 | 0.971 | 0.976 | 0.995 | 0.992 | 0.981 | 0.975 | 0.976 | 0.986 | 0.978 |
| | | 0.7 | 0.3 | 0.973 | 0.994 | 0.999 | 0.992 | 0.991 | 0.988 | 0.997 | 0.988 | 0.999 | 0.986 | 0.980 | 0.992 |
| | | 0.7 | 0.7 | 0.982 | 0.993 | 0.982 | 0.999 | 0.979 | 0.976 | 0.997 | 0.988 | 0.996 | 0.987 | 0.993 | 0.991 |
| | 4 | 0.3 | 0.3 | 0.974 | 0.999 | 0.990 | 0.982 | 0.994 | 0.970 | 0.998 | 0.988 | 0.994 | 0.976 | 1.000 | 0.993 |
| | | 0.3 | 0.7 | 0.977 | 0.986 | 0.991 | 0.990 | 0.997 | 0.981 | 0.984 | 0.997 | 0.986 | 0.987 | 0.972 | 0.975 |
| | | 0.7 | 0.3 | 0.999 | 0.976 | 0.986 | 0.972 | 0.976 | 0.986 | 0.998 | 0.995 | 0.991 | 0.977 | 0.971 | 0.979 |
| | | 0.7 | 0.7 | 0.973 | 0.996 | 0.997 | 0.977 | 0.988 | 0.997 | 0.994 | 0.990 | 0.978 | 0.988 | 0.988 | 0.984 |
| Frank | 1 | 0.3 | 0.3 | 0.990 | 0.981 | 0.979 | 0.998 | 0.987 | 0.993 | 0.973 | 0.983 | 0.999 | 0.976 | 0.979 | 0.989 |
| | | 0.3 | 0.7 | 0.984 | 0.989 | 0.990 | 0.985 | 0.990 | 0.999 | 0.978 | 0.987 | 0.980 | 0.987 | 0.972 | 0.982 |
| | | 0.7 | 0.3 | 0.984 | 0.996 | 0.971 | 0.979 | 0.988 | 0.998 | 0.980 | 0.988 | 0.982 | 0.984 | 0.979 | 0.991 |
| | | 0.7 | 0.7 | 0.992 | 0.977 | 0.986 | 0.992 | 0.997 | 0.996 | 0.991 | 0.974 | 0.972 | 0.987 | 0.972 | 0.996 |
| | 2 | 0.3 | 0.3 | 0.986 | 0.992 | 0.974 | 0.989 | 0.973 | 0.982 | 0.997 | 0.976 | 0.979 | 0.981 | 0.978 | 0.971 |
| | | 0.3 | 0.7 | 0.993 | 0.971 | 0.982 | 0.977 | 0.991 | 0.981 | 0.980 | 0.988 | 0.983 | 0.991 | 0.973 | 0.987 |
| | | 0.7 | 0.3 | 0.973 | 0.982 | 0.989 | 0.985 | 0.975 | 0.986 | 0.983 | 0.998 | 0.982 | 0.976 | 0.992 | 0.972 |
| | | 0.7 | 0.7 | 0.984 | 0.974 | 0.988 | 0.980 | 0.978 | 0.989 | 0.981 | 0.983 | 0.987 | 0.997 | 0.997 | 0.979 |
| | 3 | 0.3 | 0.3 | 0.970 | 0.987 | 0.970 | 0.979 | 0.995 | 0.992 | 0.979 | 0.994 | 0.988 | 0.986 | 0.999 | 0.997 |
| | | 0.3 | 0.7 | 0.976 | 0.995 | 0.976 | 0.999 | 0.986 | 0.994 | 0.980 | 0.996 | 0.988 | 0.984 | 0.982 | 0.984 |
| | | 0.7 | 0.3 | 0.996 | 0.992 | 0.987 | 0.976 | 0.979 | 0.984 | 0.982 | 0.990 | 0.998 | 0.992 | 0.993 | 0.990 |
| | | 0.7 | 0.7 | 0.987 | 0.990 | 0.975 | 0.974 | 0.998 | 0.978 | 0.988 | 0.972 | 1.000 | 0.994 | 0.987 | 0.973 |
| | 4 | 0.3 | 0.3 | 0.997 | 0.981 | 0.994 | 0.993 | 0.985 | 0.979 | 0.986 | 0.991 | 0.985 | 0.996 | 0.984 | 0.987 |
| | | 0.3 | 0.7 | 0.975 | 0.987 | 0.980 | 0.978 | 0.971 | 0.987 | 0.998 | 0.990 | 0.997 | 0.992 | 0.985 | 0.993 |
| | | 0.7 | 0.3 | 0.987 | 0.975 | 0.973 | 0.995 | 0.981 | 0.973 | 0.993 | 0.970 | 0.978 | 0.983 | 0.982 | 0.995 |
| | | 0.7 | 0.7 | 0.984 | 0.987 | 0.998 | 0.999 | 0.982 | 0.996 | 0.983 | 0.988 | 0.996 | 0.992 | 0.986 | 0.991 |

## 4. Conclusion

The analysis of hydrological and climatological datasets often requires the detection of relationships between different variables. When we face with the relationship of two time series, cross-correlation function and copula models are suggested. Cross-correlation function is somewhat sensitive to abnormality of datasets and existence of outliers. An efficient way to model dependency is to use new modeling mechanism of Copula Theory which helps understand the correlation beyond linearity. Copula technique is most efficient for stationary time series and may not work well for non-stationary time series such as cyclostationary time series. To solve this issue, in this research, we introduced a copula-based regression technique. The ability of the proposed approach to detect relationship between two cyclostationary time series with fractional Brownian motion errors was studied. For this purpose, numerous datasets from two cyclostationary time series with fractional Brownian motion errors were produced and analyzed. The results indicated that the values of goodness of fit indices were close to one, and consequently, the proposed method was robust to detect relationship between two cyclostationary time series with fractional Brownian motion errors.

**References**

[1] Li, Q., He, P., He, Y., Han, X., Zeng, T., Lu, G., & Wang, H. (2020). Investigation to the relation between meteorological drought and hydrological drought in the upper Shaying River Basin using wavelet analysis. *Atmospheric Research*, *234*, 104743.

[2] Petty, T. R., & Dhingra, P. (2018). Streamflow hydrology estimate using machine learning (SHEM). *JAWRA Journal of the American Water Resources Association*, *54*(1), 55-68.

[3] Liang, J., Meng, Q., Li, X., Yuan, Y., Peng, Y., Li, X., ... & Yan, M. (2021). The influence of hydrological variables, climatic variables and food availability on Anatidae in interconnected river-lake systems, the middle and lower reaches of the Yangtze River floodplain. *Science of the Total Environment*, *768*, 144534.

[4] Peña-Gallardo, M., Vicente-Serrano, S. M., Hannaford, J., Lorenzo-Lacruz, J., Svoboda, M., Domínguez-Castro, F., ... & El Kenawy, A. (2019). Complex influences of meteorological drought time-scales on hydrological droughts in natural basins of the contiguous Unites States. *Journal of Hydrology*, *568*, 611-625.

[5] Mirzaee, S., Yousefi, S., Keesstra, S., Pourghasemi, H. R., Cerdà, A., & Fuller, I. C. (2018). Effects of hydrological events on morphological evolution of a fluvial system. *Journal of Hydrology*, *563*, 33-42.

[6] Ablat, X., Liu, G., Liu, Q., & Huang, C. (2019). Application of Landsat derived indices and hydrological alteration matrices to quantify the response of floodplain wetlands to river hydrology in arid regions based on different dam operation strategies. *Science of the total environment*, *688*, 1389-1404.

[7] Konapala, G., Kao, S. C., & Addor, N. (2020). Exploring Hydrologic Model Process Connectivity at the Continental Scale Through an Information Theory Approach. *Water Resources Research*, *56*(10), e2020WR027340.

[8] Xu, Y., Zhang, X., Wang, X., Hao, Z., Singh, V. P., & Hao, F. (2019). Propagation from meteorological drought to hydrological drought under the impact of human activities: A case study in northern China. *Journal of Hydrology*, *579*, 124147.

[9] Juez, C., Peña-Angulo, D., Khorchani, M., Regüés, D., & Nadal-Romero, E. (2021). 20-years of hindsight into hydrological dynamics of a mountain forest catchment in the Central Spanish Pyrenees. *Science of The Total Environment*, *766*, 142610.

[10] Tai, X., Anderegg, W. R., Blanken, P. D., Burns, S. P., Christensen, L., & Brooks, P. D. (2020). Hillslope hydrology influences the spatial and temporal patterns of remotely sensed ecosystem productivity. *Water Resources Research*, *56*(11), e2020WR027630.

[11] Mallick, J., Talukdar, S., Alsubih, M., Salam, R., Ahmed, M., Kahla, N. B., & Shamimuzzaman, M. (2021). Analysing the trend of rainfall in Asir region of Saudi Arabia using the family of Mann-Kendall tests, innovative trend analysis, and detrended fluctuation analysis. *Theoretical and Applied Climatology*, *143*(1), 823-841.

[12] Chen, J., Li, C., Brissette, F. P., Chen, H., Wang, M., & Essou, G. R. (2018). Impacts of correcting the inter-variable correlation of climate model outputs on hydrological modeling. *Journal of hydrology*, *560*, 326-341.

[13] Myronidis, D., Ioannou, K., Fotakis, D., & Dörflinger, G. (2018). Streamflow and hydrological drought trend analysis and forecasting in Cyprus. *Water resources management*, *32*(5), 1759-1776.

[14] Abeysingha, N. S., Wickramasuriya, M. G., & Meegastenna, T. J. (2020). Assessment of meteorological and hydrological drought; a case study in Kirindi Oya river basin in Sri Lanka. *International Journal of Hydrology Science and Technology*, *10*(5), 429-447.

[15] Atif, I., Iqbal, J., & Mahboob, M. A. (2018). Investigating Snow Cover and Hydrometeorological Trends in Contrasting Hydrological Regimes of the Upper Indus Basin. *Atmosphere*, *9*(5), 162.

[16] Wang, S., Zhang, K., van Beek, L. P., Tian, X., & Bogaard, T. A. (2020). Physically-based landslide prediction over a large region: Scaling low-resolution hydrological model results for high-resolution slope stability assessment. *Environmental Modeling & Software*, *124*, 104607.

[17] Myronidis, D., Ioannou, K., Fotakis, D., & Dörflinger, G. (2018). Streamflow and hydrological drought trend analysis and forecasting in Cyprus. *Water resources management*, *32*(5), 1759-1776.

[18] Seo, S. B., Bhowmik, R. D., Sankarasubramanian, A., Mahinthakumar, G., & Kumar, M. (2019). The role of cross-correlation between precipitation and temperature in basin-scale simulations of hydrologic variables. *Journal of Hydrology*, *570*, 304-314.

[19] Li, Q., He, P., He, Y., Han, X., Zeng, T., Lu, G., & Wang, H. (2020). Investigation to the relation between meteorological drought and hydrological drought in the upper Shaying River Basin using wavelet analysis. *Atmospheric Research*, *234*, 104743.

[20] Dong, J., Wei, L., Chen, X., Duan, Z., & Lu, Y. (2020). An instrument variable based algorithm for estimating cross-correlated hydrological remote sensing errors. *Journal of Hydrology*, *581*, 124413.

[21] Liu, Y. R., Li, Y. P., Ma, Y., Jia, Q. M., & Su, Y. Y. (2020). Development of a Bayesian-copula-based frequency analysis method for hydrological risk assessment–The Naryn River in Central Asia. *Journal of Hydrology*, *580*, 124349.

[22] Ni, L., Wang, D., Wu, J., Wang, Y., Tao, Y., Zhang, J., ... & Xie, F. (2020). Vine copula selection using mutual information for hydrological dependence modeling. Environmental research, 186, 109604.

[23] Poonia, V., Jha, S., & Goyal, M. K. (2021). Copula based analysis of meteorological, hydrological and agricultural drought characteristics across Indian river basins. International Journal of Climatology.

[24] Tahroudi, M. N., Ramezani, Y., De Michele, C., & Mirabbasi, R. (2020). A new method for joint frequency analysis of modified precipitation anomaly percentage and streamflow drought index based on the conditional density of copula functions. Water Resources Management, 34(13), 4217-4231.

[25] Fan, Y., Huang, K., Huang, G. H., & Li, Y. P. (2020). A factorial Bayesian copula framework for partitioning uncertainties in multivariate risk inference. Environmental research, 183, 109215.

[26] Wang, F., Wang, Z., Yang, H., Di, D., Zhao, Y., & Liang, Q. (2020). A new copula-based standardized precipitation evapotranspiration streamflow index for drought monitoring. Journal of Hydrology, 585, 124793.

[27] Sklar, M. (1959). Fonctions de repartition an dimensions et leurs marges. Publ. inst. statist. univ. Paris, 8, 229-231.

[28] Nelson, R. B. (2006). An Introduction to Copulas. Springer New York.

[29] Schweizer, B. (1991). Thirty years of copulas. In Advances in probability distributions with given marginals (pp. 13-50). Springer, Dordrecht.

[30] Masarotto, G., & Varin, C. (2012). Gaussian copula marginal regression. Electronic Journal of Statistics, 6, 1517-1549.

[31] Gumbel, E. J. (1960). Copula Distribution des valeurs extremes en plusieurs dimensions in finance. Publications de I'Institut de Statistique de I'Universite de Paris, 9, 171–173.

[32] Clayton, D. G. (1978). A Model for Association in Bivariate Life Tables and Its Application in Epidemiological Studies of Familial Tendency in Chronic Disease Incidence. Biometrika, 65(1), 141.

[33] Demarta, S., & McNeil, A. J. (2005). The t copula and related copulas. International statistical review, 73(1), 111-129.

[34] Frank, M. J. (1979). On the Simultaneous Associativity of F (x, y) and x+y-F(x,y). Equtiones Mathematice. Retrieved from https://eudml.org/doc/136825.

[35] Nematollahi, A. R., Soltani, A. R., & Mahmoudi, M. R. (2017). Periodically correlated modeling by means of the periodograms asymptotic distributions. Statistical Papers, 58(4), 1267-1278.

[36] Mahmoudi, M. R., & Maleki, M. (2017). A new method to detect periodically correlated structure. Computational Statistics, 32(4), 1569-1581.

[37] Heydari, M. H., Mahmoudi, M. R., Shakiba, A., Avazzadeh, Z. (2018). Chebyshev cardinal wavelets and their application in solving nonlinear stochastic differential equations with fractional Brownian motion. Communications in Nonlinear Science and Numerical Simulation, 64, 98-121.

[38] Heydari, M. H., Avazzadeh, Z., Mahmoudi, M. R. (2019). Chebyshev cardinal wavelets for nonlinear stochastic differential equations driven with variable-order fractional Brownian motion. Chaos, Solitons & Fractals 124, 105-124.